\documentclass[eqsecnum,twocolumn,aps,epsf]{revtex4}
\usepackage{graphicx}
\usepackage{color}
\begin{document}
\draft

\title{Development of Cu-spin correlation in Bi$_{1.74}$Pb$_{0.38}$Sr$_{1.88}$Cu$_{1-y}$Zn$_y$O$_{6+\delta}$ high-temperature superconductors observed by muon spin relaxation}

\author{T. Adachi}
\thanks{Corresponding author: adachi@teion.apph.tohoku.ac.jp}
\author{Y. Tanabe}
\altaffiliation[Present address: ]{WPI-Advanced Institute of Materials Research, Tohoku University, 6-3 Aoba, Aramaki, Aoba-ku, Sendai 980-8579, Japan}
\author{K. Suzuki}
\author{Y. Koike}
\affiliation{Department of Applied Physics, Graduate School of Engineering, Tohoku University, 6-6-05 Aoba, Aramaki, Aoba-ku, Sendai 980-8579, Japan}

\author{T. Suzuki}
\altaffiliation[Present address: ]{College of Liberal Arts, International Christian University, 3-10-2 Osawa, Mitaka, Tokyo 181-8585, Japan}
\author{T. Kawamata}
\altaffiliation[Present address: ]{Department of Applied Physics, Graduate School of Engineering, Tohoku University, 6-6-05 Aoba, Aramaki, Aoba-ku, Sendai 980-8579, Japan}
\author{I. Watanabe}
\affiliation{Advanced Meson Science Laboratory, Nishina Center for Accelerator-Based Science, RIKEN, 2-1 Hirosawa, Wako 351-0198, Japan}

\date{\today}

\begin{abstract}

A systematic muon-spin-relaxation study in Bi-2201 high-$T_{\rm c}$ cuprates has revealed that the Cu-spin correlation (CSC) is developed at low temperatures below 2 K in a wide range of hole concentration where superconductivity appears. 
The CSC tends to become weak gradually with increasing hole-concentration. 
Moreover, CSC has been enhanced through the 3 \% substitution of Zn for Cu. 
These results are quite similar to those observed in La-214 high-$T_{\rm c}$ cuprates. 
Accordingly, it has been suggested that the intimate relation between the so-called spin-charge stripe correlations and superconductivity is a universal feature in hole-doped high-$T_{\rm c}$ cuprates.
Furthermore, apparent development of CSC, which is suppressed through the Zn substitution oppositely, has been observed in non-superconducting heavily overdoped samples, being argued in the context of a recently proposed ferromagnetic state in heavily overdoped cuprates. 

\end{abstract}
\vspace*{2em}
\pacs{PACS numbers: 74.25.Ha,76.75.+i,74.62.Dh,74.72.Gh}
\maketitle
\newpage

\section{Introduction}
Interrelation between magnetism and superconductivity continues to be a subject of extensive study in the research of the high-$T_{\rm c}$ cuprate superconductivity. 
In the underdoped regime of La-214 and Y-123 cuprates, muon spin relaxation ($\mu$SR)~\cite{kitazawa,niedermayer,savici} and nuclear magnetic/quadrapole resonance (NMR/NQR)~\cite{nabe,nabe-2,alloul,julien} experiments have uncovered the emergence of a spin-glass-like static magnetic state below the superconducting transition temperature, $T_{\rm c}$.
Moreover, an incommensurate static magnetic state relating to the so-called spin-charge stripe order~\cite{tranquada} has been observed from neutron-scattering experiments in underdoped La-214 cuprates.~\cite{suzuki,fujita} 
In the overdoped regime of La$_{2-x}$Sr$_x$CuO$_4$ (LSCO), a linear relationship between the dynamic spin susceptibility due to the incommensurate Cu-spin correlation (CSC) and $T_{\rm c}$ has been suggested from neutron-scattering experiments.~\cite{wakimoto}
These results indicate the existence of CSC in a wide range of hole concentration where superconductivity appears, and therefore imply a significant relationship between CSC and high-$T_{\rm c}$ superconductivity.

It is well-known that CSC tends to be developed through the partial substitution of impurities for Cu in high-$T_{\rm c}$ cuprates. 
Our comprehensive $\mu$SR studies on partially Zn-substituted LSCO~\cite{watanabe,adachi,risdi,adachi2} have revealed that CSC is developed through the substitution of a small amount of Zn in the whole superconducting regime of $0.05 < x < 0.28$ and that a magnetic order is formed at low temperatures below $x=0.15$. 
These can be understood in terms of the Zn-induced development of the stripe correlations, suggesting an intimate relation between the stripe correlations and superconductivity. 

One may doubt, however, the universality of the superconductivity relating to the stripes, since effects of the stripes on various properties appear markedly in the La-214 cuprates. 
Here, it is the mono-layer Bi-2201 cuprate that has a potential to clear off the above suspicion, because the partial substitution of Pb or La in the Bi-2201 cuprate allows us to have samples with a wide range of hole concentration from the underdoped to overdoped regime.~\cite{ando,kudo}
Russo {\it et al}.~\cite{russo} and Miyazaki {\it et al}.~\cite{miyazaki} have reported from $\mu$SR measurements in Bi$_2$Sr$_{2-x}$La$_x$CuO$_{6+\delta}$ and Bi$_{1.76}$Pb$_{0.35}$Sr$_{1.89}$CuO$_{6+\delta}$, respectively, that the development of CSC is absent at low temperatures down to 2 K in the superconducting regime.
As for the Zn-substituted Bi-2212 cuprate, on the other hand, the development of CSC has been observed at very low temperatures below 2 K around the hole concentration per Cu, $p$, of 1/8,~\cite{akoshima,akoshima2} implying that the development of CSC is weak in the Bi-based cuprates compared with the La-214 cuprates. 

In this paper, $\mu$SR study of CSC in the partially Pb-substituted Bi-2201 cuprate has been reported. 
It has been observed that the depolarization of muon spins due to the development of CSC becomes marked with decreasing temperature below 2 K and that the Zn substitution enhances CSC in the whole superconducting regime of the Bi-2201 cuprate. 
These results are in qualitative agreement with those observed in the La-214 cuprates,~\cite{watanabe,adachi,risdi,adachi2} strongly suggesting the universal relation between the stripe correlations and superconductivity in hole-doped high-$T_{\rm c}$ cuprates. 
Unexpectedly, a slight enhancement of the depolarization of muon spins has been observed in non-superconducting heavily overdoped samples. 
This may be consistent with the recent suggestion in LSCO that a ferromagnetic state appears in the non-superconducting heavily overdoped regime.~\cite{sonier2}

\section{Experimental}
Samples of Bi$_{1.74}$Pb$_{0.38}$Sr$_{1.88}$Cu$_{1-y}$Zn$_y$O$_{6+\delta}$ (BPSCZO) polycrystals used for the present measurements were prepared by the ordinary solid-state reaction method.~\cite{kudo} 
The $p$ value was controlled by annealing the samples in vacuum (Vac) or oxygen (Oxy) for $24-48$ h. 
The characterization of the samples was performed by the x-ray diffraction, electrical resistivity, $\rho$, and magnetic susceptibility, $\chi$, measurements. 
Moreover, the composition of the samples was checked by the inductively-coupled-plasma (ICP) analysis to be stoichiometric within the experimental error. 
Zero-field (ZF) $\mu$SR measurements were performed down to 0.3 K at the RIKEN-RAL Muon Facility at the Rutherford-Appleton Laboratory in the UK, using a pulsed positive surface muon beam. 

\begin{figure}[tbp]
\begin{center}
\includegraphics[width=1.0\linewidth]{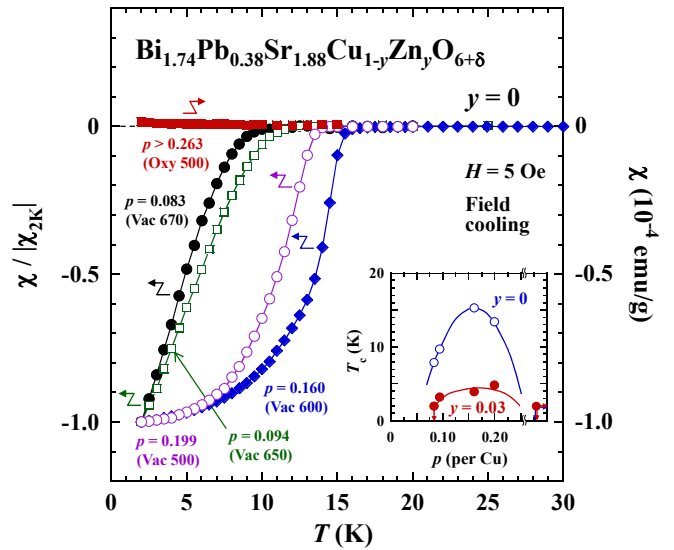}
\end{center}
\caption{(Color online) Temperature dependence of the magnetic susceptibility, $\chi$, in a magnetic field of 5 Oe on field cooling for Bi$_{1.74}$Pb$_{0.38}$Sr$_{1.88}$Cu$_{1-y}$Zn$_y$O$_{6+\delta}$ with $y=0$. Values of $\chi$ except for $p>0.263$ are normalized by the value of $\chi$ at 2 K, $|\chi_{\rm 2K}|$. The inset shows the $p$ dependence of the superconducting transition temperature, $T_{\rm c}$, in Zn-free and 3 \% Zn-substituted samples.} 
\label{fig:fig1} 
\end{figure}

Figure 1 shows the temperature dependence of $\chi$ in a magnetic field of 5 Oe on field cooling for Zn-free BPSCZO with $y=0$. 
The values of $p$ were defined using the empirical law of $T_{\rm c} / T_{\rm c}^{\rm max} = 1 - 82.6(p - 0.16)^2$ where $T_{\rm c}^{\rm max}$ is the maximum value of $T_{\rm c}$ in a system,~\cite{presland} although how to estimate the $p$ value is an unsettled issue in the Bi-2201 cuprate.~\cite{ando,kudo2,comment} 
The $T_{\rm c}$ increases with decreasing annealing-temperature in vacuum, namely, with increasing $p$ and shows the maximum at 600$^{\rm o}$C, followed by the decrease in $T_{\rm c}$ and the disappearance of superconductivity for the oxygenated sample, as also shown in the inset. 
As for Zn-substituted BPSCZO, on the other hand, it is found that $T_{\rm c}$ is apparently reduced for $y=0.03$ in the whole superconducting regime, as shown in the inset. 
Together with the ICP result, it is concluded that Zn is expectedly substituted for Cu in BPSCZO.

\section{Results}
\begin{figure}[tbp]
\begin{center}
\includegraphics[width=1.0\linewidth]{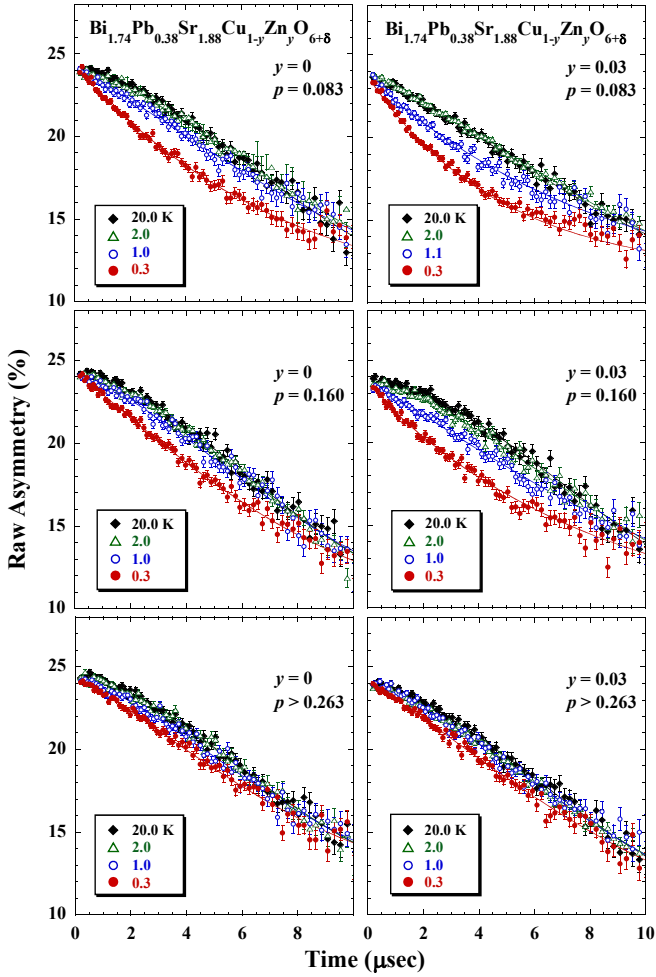}
\end{center}
\caption{(Color online) Zero-field $\mu$SR time spectra at various temperatures for typical values of the hole concentration, $p$, in Bi$_{1.74}$Pb$_{0.38}$Sr$_{1.88}$Cu$_{1-y}$Zn$_y$O$_{6+\delta}$ with $y=0$ and 0.03.} 
\label{fig:fig2} 
\end{figure}

Figure 2 shows the ZF-$\mu$SR time spectra of the underdoped ($p=0.083$), optimally doped ($p=0.160$), non-superconducting heavily overdoped ($p>0.263$) samples of BPSCZO with $y=0$ and 0.03. 
For the Zn-free sample of $p=0.083$, the spectra above 2 K show Gaussian-like depolarization due to randomly oriented nuclear spins, indicating that the Cu spins fluctuate on a much shorter time scale than the $\mu$SR time window of $10^{-11} - 10^{-6}$ s. 
Below 2 K, on the other hand, it is found that the depolarization of muon spins is enhanced and an exponential-like depolarization is observed around 0.3 K, suggesting the development of CSC at very low temperatures below 2 K. 
For the 3 \% Zn-substituted samples with $y=0.03$, it is found that the depolarization below 2 K is larger than that for the Zn-free samples with $y=0$, suggesting the Zn-induced enhancement of CSC in BPSCZO as well as in Zn-substituted LSCO.~\cite{watanabe,adachi,risdi,adachi2}

With increasing $p$ both in $y=0$ and 0.03, the depolarization at low temperatures becomes weak, as seen in the spectra of $p=0.160$. 
Further increase in $p$ results in further weakening of the depolarization in the overdoped sample and no exponential-like depolarization is observed for the non-superconducting samples of $p>0.263$ both with $y=0$ and 0.03. 
These results suggest the weakening of CSC with doping of holes. 

Surprisingly, it is found both in $y=0$ and 0.03 that the spectra at 0.3 K for $p>0.263$ exhibit a slight but apparent change from the Gaussian-like depolarization. 
For Zn-substituted LSCO, it has been reported that the Zn-induced development of CSC appears within the superconducting regime.~\cite{risdi} 
The present results indicate that CSC is developed even in the non-superconducting heavily overdoped sample of BPSCZO.

To give a further insight into CSC, the spectra were analyzed using the following standard function: $A(t) = A_0 {\rm exp}(-\lambda_0 t) G_{\rm Z} (\Delta,t) + A_1 {\rm exp}(-\lambda_1 t) + A_{\rm BG}$. 
The first term represents the slowly depolarizing component in a region where the Cu spins fluctuate fast beyond the $\mu$SR time window. 
The $G_{\rm Z}(\Delta,t)$ is the static Kubo-Toyabe function with a half width of the static internal field at the muon site, $\Delta$, describing the distribution of the nuclear-dipole field at the muon site. 
The second term represents the fast depolarizing component in a region where the Cu-spin fluctuations slow down. 
The $A_0$ and $A_1$ are initial asymmetries and $\lambda_0$, $\lambda_1$ are depolarization rates of each component, respectively. 
The $A_{\rm BG}$ is the temperature-independent background term.
All the spectra except for those of $p \le 0.160$ with $y=0.03$ around 0.3 K are well reproduced by only the first and third terms of the above function. 
This is contrary to the results of LSCO in which the depolarization of muon spins is dramatically fast and muon-spin precession is observed at low temperatures, indicating that the development of CSC in BPSCZO is weaker than in LSCO. 

\begin{figure}[tbp]
\begin{center}
\includegraphics[width=1.0\linewidth]{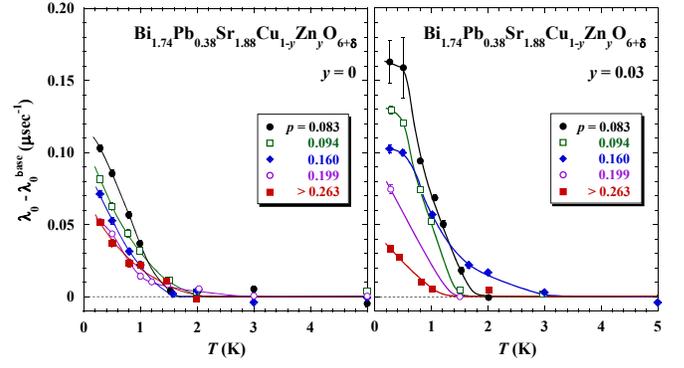}
\end{center}
\caption{(Color online) Temperature dependence of the depolarization rate of muon spins, $\lambda_0$, subtracted by the temperature-independent depolarization rate above 3 K, $\lambda_0^{\rm base}$, for Bi$_{1.74}$Pb$_{0.38}$Sr$_{1.88}$Cu$_{1-y}$Zn$_y$O$_{6+\delta}$ with $y=0$ and 0.03. Solid lines are to guide the reader's eye.} 
\label{fig:fig3} 
\end{figure}

Figure 3 shows the temperature dependence of $\lambda_0$ subtracted by the temperature-independent depolarization rate above 3 K, $\lambda_0^{\rm base}$, for $y=0$ and 0.03. 
Below $\sim 3$ K, $\lambda_0 - \lambda_0^{\rm base}$ tends to increase gradually with decreasing temperature both for $y=0$ and 0.03. 
For $y=0$, $\lambda_0 - \lambda_0^{\rm base}$ at 0.3 K decreases with increasing $p$ but is finite even in the non-superconducting sample of $p>0.263$. 
For $y=0.03$, on the other hand, the enhancement of $\lambda_0 - \lambda_0^{\rm base}$ is marked at low temperatures for $p=0.083$ and tends to be saturated below 0.5 K, which is a typical behavior toward a static magnetic order.
At 0.3 K, $\lambda_0 - \lambda_0^{\rm base}$ decreases with increasing $p$ but is finite for $p>0.263$. 
Intriguingly, $\lambda_0 - \lambda_0^{\rm base}$ at 0.3 K is enhanced through the Zn substitution in the superconducting regime, while it is reduced in the non-superconducting regime. 

Temperature dependence of $\Delta$ for $y=0$ and 0.03 is shown in Fig. 4. 
Both for $y=0$ and 0.03, $\Delta$ is almost constant above $\sim 2$ K, while it tends to decrease with decreasing temperature below $\sim 2$ K where $\lambda_0$ tends to increase. 
It is found that the decrease in $\Delta$ below $\sim 2$ K is marked for samples exhibiting a notable increase in $\lambda_0 - \lambda_0^{\rm base}$ as shown in Fig. 3 and, moreover, $\Delta$ reaches zero at 0.3 K for $p \le 0.160$ with $y = 0.03$.
In general, $\Delta$ is finite and independent of temperature owing to effects of nuclear-dipole field being unchanged at low temperatures.
The decrease in $\Delta$ below $\sim 2$ K is regarded as owing to the development of CSC. 
That is, the effect of the nuclear-dipole field tends to be masked by the effetc of Cu spins due to the development of CSC, which was usually observed in LSCO.~\cite{nabe-gaussian}
It is noted that both changes of $\Delta$ and $\lambda_0$ below $\sim 2$ K are not due to a trade between them, because $\Delta$ reflecting a static field and $\lambda_0$ reflecting a dynamic field are parameters distinct from each other.

The $p$ dependence of $\lambda_0 - \lambda_0^{\rm base}$ at 0.3 K for BPSCZO with $y=0$ and 0.03 is displayed in Fig. 5, together with that of $T_{\rm c}$. 
It is found that $\lambda_0 - \lambda_0^{\rm base}$ is finite and is enhanced through the Zn substitution for all samples in the superconducting regime. 
Moreover, $\lambda_0 - \lambda_0^{\rm base}$ tends to decrease monotonically with an increase in $p$ both for $y=0$ and 0.03. 
These suggest a possible relation between the development of CSC and superconductivity in BPSCZO. 
Another attention should be paid to the reduction of $\lambda_0 - \lambda_0^{\rm base}$ through the Zn substitution for $p>0.263$, implying that the mechanism of CSC is different between the superconducting and non-superconducting heavily overdoped regimes.

\section{Discussion}
\begin{figure}[tbp]
\begin{center}
\includegraphics[width=1.0\linewidth]{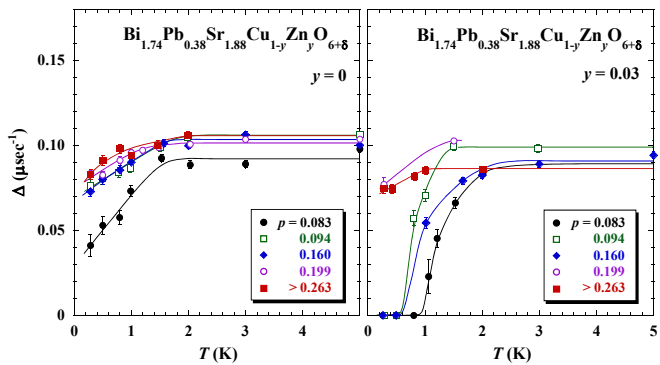}
\end{center}
\caption{(Color online) Temperature dependence of a half width of the static internal field at the muon site, $\Delta$, for Bi$_{1.74}$Pb$_{0.38}$Sr$_{1.88}$Cu$_{1-y}$Zn$_y$O$_{6+\delta}$ with $y=0$ and 0.03. Solid lines are to guide the reader's eye.} 
\label{fig:fig5} 
\end{figure}

The present $\mu$SR results have provided following significant information in the Bi-2201 cuprate. 
One is that CSC is developed at low temperatures in the whole superconducting regime. 
An early NMR experiment has revealed the formation of a magnetic order in the parent compound of the Bi-2201 cuprate, Bi$_2$Sr$_2$CuO$_{6+\delta}$.~\cite{kato} 
In the superconducting regime, on the contrary, no clear evidence for the development of CSC has been reported from $\mu$SR measurements, as described before.~\cite{russo,miyazaki}
Therefore, the present result is the {\it first} evidence for the development of CSC in the superconducting regime of the Bi-2201 cuprate, attained by lowering the measurement temperature below 2 K. 

The other characteristic is the Zn-induced development of CSC in the superconducting regime. 
A recent inelastic neutron-scattering experiment has found incommensurate magnetic peaks at low temperatures in the underdoped regime of Bi$_{2+x}$Sr$_{2-x}$CuO$_{6+\delta}$.~\cite{enoki}
In an elastic neutron-scattering experiment, moreover, incommensurate static magnetic peaks have been observed for the Fe-substituted Bi$_{1.75}$Pb$_{0.35}$Sr$_{1.90}$Cu$_{1-y}$Fe$_y$O$_{6+\delta}$ in the overdoped regime.~\cite{hiraka} 
These neutron-scattering results are very similar to those observed in LSCO,~\cite{yamada,fujita2} suggesting that the observed CSC in BPSCZO is the stripe correlations in origin. 
As observed in the present $\mu$SR and neutron-scattering results,~\cite{enoki,enoki2} the unambiguous characteristic in BPSCZO is that the development of CSC is not so noticeable as in LSCO. 
This may be due to the weakness of the stripe correlations and/or the strong two-dimensionality of the crystal structure in BPSCZO due to the long distance between the neighboring CuO$_2$ planes. 
Accordingly, it is strongly suggested that CSC in BPSCZO originates from the stripe correlations and that there exists the intimate relation between the stripe correlations and superconductivity. 
Combined with the results in LSCO, it is much convinced that the stripe correlations are inevitable for the appearance of the high-$T_{\rm c}$ superconductivity. 

A distinct feature found in the Zn-free samples of BPSCZO is the development of CSC in the overdoped regime, while it is absent in the Zn-free overdoped LSCO.~\cite{risdi,sonier2} 
Moreover, no static magnetic peaks have been observed to date in the Zn-free overdoped LSCO~\cite{wakimoto} and Bi-2201~\cite{hiraka} from any neutron-scattering experiments. 
Recent transverse-field $\mu$SR measurements in the overdoped LSCO have suggested the appearance of a field-induced heterogeneous magnetic state distributed in a form of islands around dopant Sr ions.~\cite{macd}
Based on this suggestion, the present observation of CSC in the Zn-free overdoped BPSCZO might be regarded as a fluctuation of the possible field-induced hetero-magnetism.

The observation of the slight enhancement of the depolarization rate in the non-superconducting heavily overdoped sample of BPSCZO seems to be a challenge to the common wisdom and might be related to the recent suggestion in LSCO on the appearance of a ferromagnetic state. 
Sonier {\it et al}. have observed from ZF-$\mu$SR measurements in the non-superconducting heavily overdoped LSCO with $x=0.33$ that the depolarization rate is enhanced at very low temperatures below 0.9 K.~\cite{sonier2}
Considering the $\rho$ and $\chi$ results, they have proposed the formation of a ferromagnetic order about Sr-rich clusters. 
The slightly enhanced depolarization of muon spins below 2 K in the non-superconducting heavily overdoped sample of BPSCZO might have the same origin as in the case of LSCO, although further investigation should be performed to pin down this issue. 
Intriguingly, the depolarization rate is reduced through the Zn substitution in the non-superconducting heavily overdoped sample as shown in Fig. 5, which is opposite to that observed in the superconducting regime. 
This implies a different origin of the development of CSC from the stripe-related magnetism in the superconducting regime. 
Assuming the formation of a ferromagnetic state, it may be locally destroyed around Zn, leading to the decrease in the depolarization rate through the Zn substitution.

\begin{figure}[tbp]
\begin{center}
\includegraphics[width=1.0\linewidth]{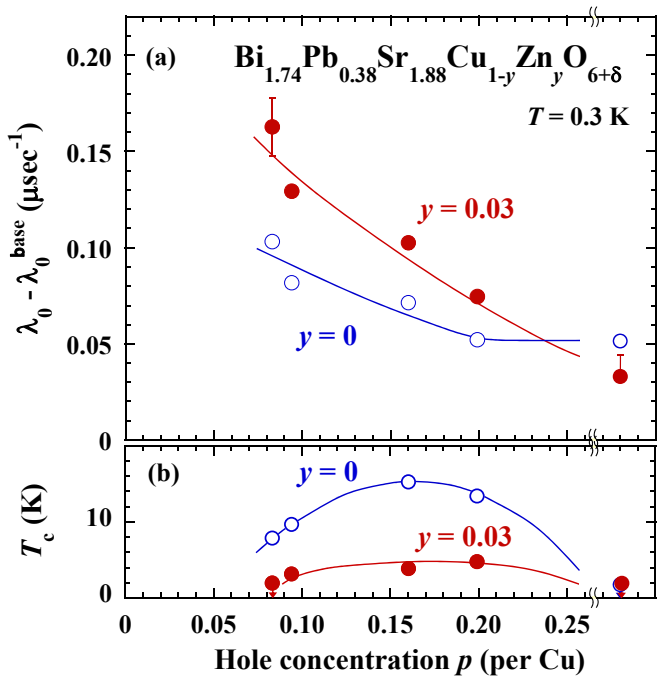}
\end{center}
\caption{(Color online) Hole-concentration, $p$, dependences of (a) the depolarization rate of muon spins subtracted by the temperature-independent depolarization rate above 3 K, $\lambda_0 - \lambda_0^{\rm base}$, and (b) $T_{\rm c}$ estimated from $\chi$ measurements for Bi$_{1.74}$Pb$_{0.38}$Sr$_{1.88}$Cu$_{1-y}$Zn$_y$O$_{6+\delta}$ with $y=0$ and 0.03. For most data, error bars are too small to be behind the symbols. Solid lines are to guide the reader's eye.} 
\label{fig:fig4} 
\end{figure}

\section{Summary}
In summary, ZF-$\mu$SR measurements of Zn-free and 3 \% Zn-substituted BPSCZO in a wide range of $p$ have revealed in the Bi-2201 cuprate that CSC is developed at low temperatures below 2 K and tends to become weak gradually with increasing $p$ in the whole superconducting regime. 
Moreover, the Zn substitution enhances CSC in the superconducting regime. 
These results are in qualitative agreement with those in LSCO, suggesting the universality of the intimate relation between the stripe correlations and superconductivity in hole-doped high-$T_{\rm c}$ cuprates.
Furthermore, a slight development of CSC has been observed in the non-superconducting heavily overdoped samples, concomitant with the suppression of CSC through the Zn substitution. 
This might be caused by the formation of a ferromagnetic state recently suggested in LSCO.~\cite{sonier2}

\section*{Acknowledgments}
We are indebted to M. Ishikuro for his help in the ICP analysis. 
The $\mu$SR measurements were partially supported by KEK-MSL Inter-University Program for Oversea Muon Facilities, by Global COE Program "Materials Integration (International Center of Education and Research), Tohoku University", MEXT, Japan, and also by "Education Program for Biomedical and Nano-Electronics, Tohoku University", Program for Enhancing Systematic Education in Graduate Schools, MEXT, Japan.


\begin{references}
\bibitem{comment}
Nevertheless, the present results insist without doubtful that CSC is developed and further enhanced through the Zn substitution in the superconducting regime and that there exists a slight but apparent development of CSC in the non-superconducting heavily overdoped regime.
\end{references}
\end{document}